\documentclass[11pt]{article}

\usepackage[preprint]{acl}
\usepackage{multirow}
\usepackage{times}
\usepackage{latexsym}

\usepackage[T1]{fontenc}

\usepackage[utf8]{inputenc}

\usepackage{microtype}

\usepackage{inconsolata}

\usepackage{graphicx}

\title{PPCR-IM: A System for Multi-layer DAG-based Public Policy Consequence Reasoning and Social Indicator Mapping}

\author{
  Zichen Song*, Weijia Li\\
  Lanzhou University \\
  \texttt{songzch21@lzu.edu.cn, forgetfuljre@gmail.com}
}

\begin{document}
\maketitle
\begin{abstract}
Public policy decisions are typically justified using a narrow set of headline indicators, leaving many downstream social impacts unstructured and difficult to compare across policies. We propose PPCR-IM, a system for multi-layer DAG-based consequence reasoning and social indicator mapping that addresses this gap. Given a policy description and its context, PPCR-IM uses an LLM-driven, layer-wise generator to construct a directed acyclic graph of intermediate consequences, allowing child nodes to have multiple parents to capture joint influences. A mapping module then aligns these nodes to a fixed indicator set and assigns one of three qualitative impact directions: increase, decrease, or ambiguous change. For each policy episode, the system outputs a structured record containing the DAG, indicator mappings, and three evaluation measures: an expected-indicator coverage score, a discovery rate for overlooked but relevant indicators, and a relative focus ratio comparing the system’s coverage to that of the government. PPCR-IM is available both as an online demo and as a configurable XLSX-to-JSON batch pipeline, and is publicly accessible at \url{https://n23pg3gbm5.coze.site}.
\end{abstract}

\section{Introduction}

Public policy decisions are typically justified and evaluated using a small set of headline indicators, such as GDP growth, inflation, or unemployment. A large literature on ``Beyond GDP'' and multidimensional well-being has shown that these metrics capture only a narrow slice of social outcomes and systematically miss distributional, environmental, and institutional dimensions.\footnote{See, e.g., overviews of multidimensional well-being and ``Beyond GDP'' indicator frameworks in economics and public policy.} In practice, the links between specific policy choices, intermediate mechanisms, and downstream indicators remain mostly implicit in narrative reports or ad hoc diagrams, which makes consequence reasoning hard to compare, audit, or reuse across policies. A system that can externalize these reasoning chains in a structured form, and align them with a shared indicator space, has the potential to make policy analysis more transparent and to surface impacts that fall outside the official focus of government dashboards\cite{2}.

Day-to-day policy work often centers on a few ``primary'' indicators that are declared ex ante in strategy or budget documents, while many second-order and long-term effects receive less systematic attention. Existing workflows provide little support for: (i) enumerating intermediate consequences beyond a short narrative; (ii) linking those consequences consistently to a fixed indicator taxonomy; and (iii) quantifying where government focus diverges from a broader set of plausible impacts. Theory-of-change diagrams and logic models, when available, are typically hand-crafted for a single program and not aligned with reusable indicator sets. PPCR-IM is designed to address this gap by turning qualitative consequence reasoning into a machine-readable object that explicitly connects policies, consequence chains, and indicators.

In this paper, we introduce \textbf{PPCR-IM} (Public Policy Consequence Reasoning and Indicator Mapping), an end-to-end system that turns qualitative policy consequence analysis into multi-layer DAG construction with indicator alignment. Given a policy description and context, PPCR-IM uses an LLM-driven, layer-wise generator to build a directed acyclic graph of intermediate consequences, and a mapping module to align these nodes to a fixed set of social indicators with qualitative impact directions (increase, decrease, or ambiguous change). For each policy episode, the system produces a structured, machine-readable record that links the policy text, the generated DAG, and indicator-level impact assessments. 

\begin{figure*}[t]
    \centering
    \includegraphics[width=\textwidth]{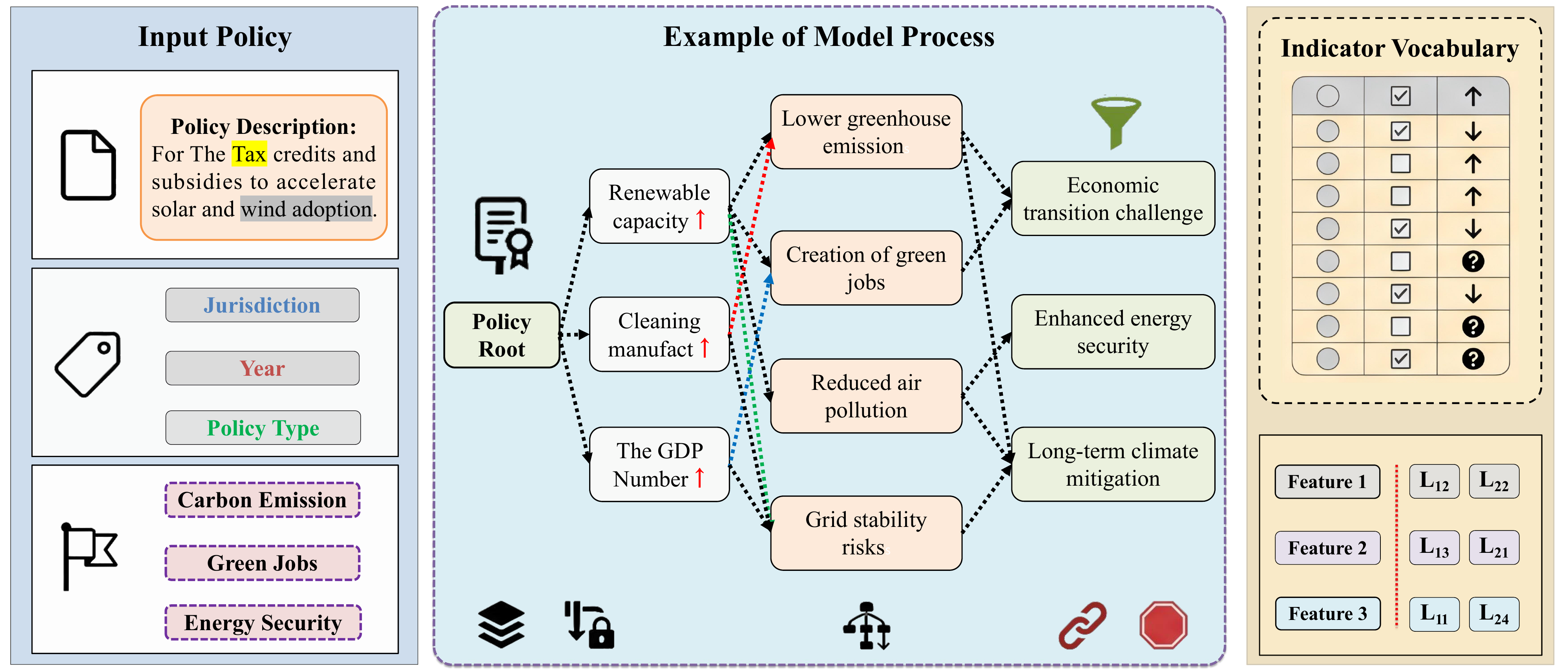}
    \caption{
    A policy description and contextual attributes are used to generate a multi-layer consequence DAG that models causal impact pathways. The system maps DAG nodes to a fixed indicator vocabulary with directional effects and supporting evidence links, enabling structured impact assessment and downstream evaluation.
    }
    \label{fig:ppcrim_overview}
\end{figure*}

Our contributions are summarized as follows:
\begin{itemize}
    \item We formulate public policy consequence reasoning as the construction of a multi-layer DAG with multi-parent dependencies, coupled with a mapping from intermediate consequences to a shared indicator space.
    \item We implement a practical system that outputs, for each policy episode, a structured record containing the DAG, indicator mappings, and evaluation measures that capture coverage of government-expected indicators, systematic discovery of overlooked but relevant indicators, and the relative focus of the system versus the government.
    \item We provide both an interactive demo and a configurable XLSX-to-JSON batch pipeline, making PPCR-IM usable by policy analysts, public administration researchers, and NLP practitioners within existing evaluation workflows.
\end{itemize}

\section{Task Formulation and Evaluation}

\subsection{Inputs and outputs}

PPCR-IM processes individual \emph{policy episodes}. Each episode is represented by a short policy description, optional contextual fields (e.g., jurisdiction, time period, macro conditions), and a set of social indicators that the government explicitly highlights in accompanying documents or expert reconstructions of those documents. In the batch setting, these fields are stored row-wise in an XLSX file so that policy analysts can prepare and edit inputs using familiar spreadsheet tools.

For each episode, the system returns a structured JSON record. This record contains: a multi-layer directed acyclic graph (DAG) of intermediate consequences; a mapping from graph nodes to a fixed indicator vocabulary, with a qualitative impact direction (increase, decrease, or ambiguous change) for each indicator; and a collection of scalar evaluation measures. These outputs are used both for interactive visualization in the demo and for downstream quantitative analysis across many policies.

\begin{figure*}[t]
    \centering
    \includegraphics[width=\textwidth]{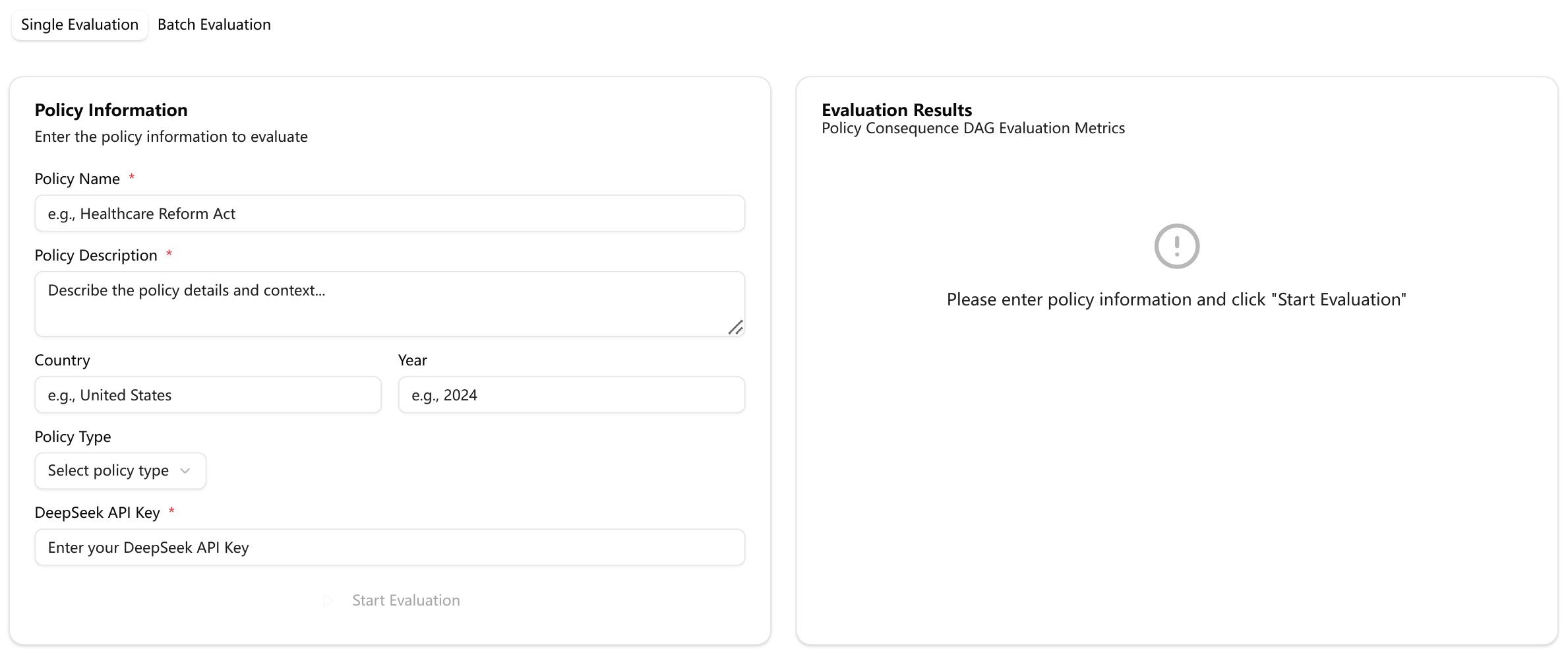}
    \caption{
    Web-based interface of the Policy Consequence DAG Evaluation System.
    Users provide policy descriptions and contextual information to trigger DAG-based consequence evaluation and indicator-level impact metrics.
    }
    \label{fig:demo_ui}
\end{figure*}

\subsection{DAG reasoning and indicator mapping}

We cast qualitative consequence reasoning as a two-step task. First, given a policy and its context, an LLM-driven, layer-wise generator constructs a DAG in which the root corresponds to the policy and subsequent layers capture downstream consequences. Nodes are short textual statements of intermediate effects, and edges encode hypothesized influence relations; child nodes may have multiple parents to represent joint influences, while depth and branching are controlled by user parameters to keep graphs interpretable. Second, PPCR-IM aligns this DAG to a fixed set of social indicators. For each indicator, the system decides whether it is plausibly affected by the policy, identifies the subset of relevant consequence nodes, and assigns a qualitative direction of change. This separation between graph construction and indicator mapping is a deliberate design choice: it allows reuse of the same indicator space across policies, supports comparison of impact patterns at the indicator level, and distinguishes PPCR-IM from approaches that only attach free-form explanations or unstructured chains of thought to individual indicators.

\subsection{Evaluation objectives}

Our evaluation focuses on how well PPCR-IM replicates and extends the indicator focus observed in government documents, rather than on intrinsic language modeling metrics or user satisfaction scores. For each policy episode with reference annotations, we compute an \emph{expected-indicator coverage score} that measures the proportion of indicators in the government focus set that the system also flags as affected by the policy. This assesses whether the system recovers the dimensions that policymakers themselves emphasize. We then assess two complementary dimensions. An \emph{overlooked-indicator discovery rate} measures how often PPCR-IM surfaces indicators that experts deem relevant but that were not highlighted by the government, capturing the system’s ability to broaden the conversation. A \emph{model–government focus ratio} compares the coverage achieved by the system and by the government on a common indicator subset, providing a normalized view of how concentrated or diffuse their respective attention is. In this work we rely on such offline, annotation-based measures; user studies and interactive evaluations are left for future work once the basic behavior of the system is well understood.

\section{System Architecture and Methods}

\subsection{Overall architecture and data flow}

Figure~\ref{fig:ppcrim_overview} summarizes the PPCR-IM architecture. The system is organized into three main modules connected by a simple data flow: a \emph{DAG generator}, an \emph{indicator mapper}, and an \emph{evaluation and export} layer. The DAG generator takes as input a policy episode record (policy text, context, and government focus indicators) and produces a multi-layer directed acyclic graph of intermediate consequences. The indicator mapper consumes this graph and produces indicator-level impact assessments. The evaluation and export layer then compares these assessments against reference annotations and writes a structured JSON record per policy episode, which is used both in the interactive demo and in the batch pipeline\cite{4}.

The overall design is intentionally modular. All communication between modules happens through explicit, machine-readable objects: the DAG generator exposes a node–edge representation with layer labels; the indicator mapper exposes a list of indicator impact entries; and the evaluation layer exposes scalar summary measures. This separation allows each module to be improved or replaced independently (for example, swapping in a different LLM backend for DAG generation, or a different indicator taxonomy), while preserving a stable interface for policy analysts. Methodologically, the architecture combines three elements that are rarely integrated in existing tools: layer-wise graph-based reasoning over policy consequences, alignment to a fixed indicator space rather than ad hoc labels, and a dedicated evaluation module that quantifies how the system’s focus compares to institutional focus across many policies.

\subsection{Layer-wise DAG construction}

The DAG generator constructs a consequence graph in a breadth-wise, layer-by-layer fashion. Starting from a root node that represents the policy itself, PPCR-IM prompts the LLM to propose downstream consequences for the current frontier of nodes, subject to user-configurable limits on maximum depth and branching factor. Each node is represented by a concise textual description and a layer index; edges are added from parents to children as the graph expands. When multiple parents propose semantically similar children, the system merges these into a single node, adds edges from all relevant parents, and reuses this node in subsequent layers. Graph expansion stops when the maximum depth is reached or when the LLM no longer proposes new, distinct consequences for a given layer. Allowing multi-parent dependencies is essential for representing realistic policy mechanisms, where many outcomes arise from the joint effect of several upstream changes. Unlike linear chain-of-thought style reasoning, which forces a single path through consequences, the DAG formulation encourages the LLM to articulate branching and converging pathways (e.g., fiscal measures and labor-market responses jointly shaping household financial stress). At the same time, depth and branching limits ensure that the resulting graphs remain compact enough for human inspection in the demo interface. This combination of constrained, layer-wise expansion and explicit multi-parent structure distinguishes PPCR-IM from systems that only attach unstructured textual rationales to model outputs.

\subsection{Indicator mapping and prediction}

Once a DAG has been constructed, PPCR-IM maps it into a fixed indicator space. The indicator mapper considers the full set of consequence nodes and a predefined vocabulary of social indicators, each with a short textual definition. For each indicator, the system queries the LLM with a structured prompt containing the policy, relevant nodes, and the indicator description, and asks whether the indicator is plausibly affected by the policy. If so, the mapper selects a subset of linked nodes and assigns a qualitative direction of change from the set \{increase, decrease, ambiguous\}. The result is a list of indicator impact entries, each tying one indicator to specific parts of the DAG and to a direction label.

This procedure serves two purposes. First, it provides a consistent way of situating heterogeneous consequence graphs in a shared indicator vocabulary, which is crucial for comparing patterns across policies and aggregating statistics over many episodes. Second, by explicitly annotating directions of change and recording which nodes support each decision, it makes it possible to define evaluation measures that distinguish between recovering government-expected indicators and discovering additional, relevant indicators that were not in the government’s declared focus\cite{14}. 


\begin{table*}[t]
\centering
\small
\begin{tabular}{llcccc}
\hline
\textbf{System} & \textbf{Metric} &
\textbf{Mean} & \textbf{Std.\ dev.} & \textbf{Min} & \textbf{Max} \\
\hline
\multirow{3}{*}{GPT~5.1} 
& Expected-indicator coverage score      & 0.851 & 0.052 & 0.68 & 0.94 \\
& Overlooked-indicator discovery rate    & 0.352 & 0.088 & 0.15 & 0.58 \\
& Model--government focus ratio          & 1.098 & 0.115 & 0.86 & 1.41 \\
\hline
\multirow{3}{*}{Doubao} 
& Expected-indicator coverage score      & 0.803 & 0.061 & 0.62 & 0.91 \\
& Overlooked-indicator discovery rate    & 0.291 & 0.079 & 0.12 & 0.51 \\
& Model--government focus ratio          & 1.023 & 0.108 & 0.81 & 1.33 \\
\hline
\multirow{3}{*}{PPCR-IM} 
& Expected-indicator coverage score      & \textbf{0.902} & 0.048 & 0.71 & 0.96 \\
& Overlooked-indicator discovery rate    & \textbf{0.603} & 0.092 & 0.18 & 0.67 \\
& Model--government focus ratio          & \textbf{1.356} & 0.124 & 0.89 & 1.52 \\
\hline
\end{tabular}
\caption{
Comparison of PPCR-IM with two LLM-based baselines on 1{,}027 policy episodes.
The coverage score measures the proportion of government-expected indicators recovered by each system.
The discovery rate quantifies the proportion of relevant but previously unfocused indicators identified.
}
\label{tab:system-comparison}
\end{table*}

\section{Experiment and Results}

\subsection{Runtime configuration and batch pipeline}

PPCR-IM is implemented as a set of Python modules with a command-line interface aimed at policy analysis teams and NLP practitioners. The main entry point takes as arguments the input XLSX file, the output directory, and runtime parameters controlling the LLM backend and graph generation process. Key options include the model identifier (\texttt{--model-name}), the generation temperature for consequence expansion (\texttt{--temperature}), a separate temperature for indicator linking (\texttt{--link-temperature}), structural constraints on the graph (\texttt{--max-depth}, \texttt{--max-branch}, \texttt{--max-links-per-node}), and API configuration (\texttt{--deepseek\_api}). In a batch run, each row of the XLSX file is treated as one policy episode, passed through the DAG generator and indicator mapper, and written out as a JSON file keyed by policy identifier\cite{3}.

For every episode, the system records a status flag indicating whether the pipeline completed successfully (\texttt{ok}), was deliberately skipped (\texttt{skipped}, e.g., due to missing mandatory fields), or encountered an error (\texttt{error}, e.g., repeated API failures or unparseable LLM output). This status, together with a short diagnostic message, is stored in the JSON alongside the DAG, indicator mappings, and evaluation measures. At the end of a run, PPCR-IM prints aggregate counts for each status type and summary statistics for the evaluation metrics, supporting both one-off experiments and repeated runs on updated datasets without modifying core code.

\subsection{Dataset and metric computation}

\begin{figure}[h]
  \centering
  \includegraphics[width=0.45\textwidth]{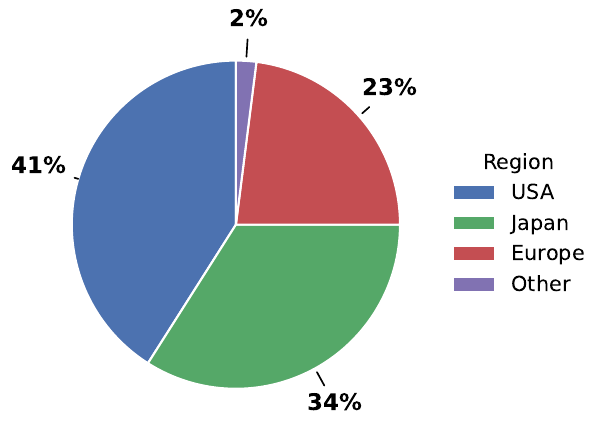}%
  \caption{Regional distribution of the 1,027 policy episodes used in our experiments.}
  \label{fig:region-pie}
\end{figure}

To evaluate PPCR-IM in a realistic setting, we construct a dataset of 1{,}027 policy episodes. We begin from a set of nineteen social and macroeconomic indicators inspired by World Bank data, covering domains such as growth, employment, debt, and social spending, and use these as the fixed indicator vocabulary\cite{5}. Each episode consists of a short policy description, basic context fields, and two reference sets: a government focus set, reflecting indicators emphasized in official communication, and a relevance set, reflecting indicators that expert annotators judge to be materially affected regardless of whether they were highlighted. The episodes span several regions; as shown in Figure~\ref{fig:region-pie}, approximately 41\% of cases come from the United States, 34\% from Japan, 23\% from European countries, and the remaining 2\% from other regions\cite{6}.

Given the JSON outputs produced by PPCR-IM on this corpus, we compute three annotation-based evaluation measures per episode: an expected-indicator coverage score (the proportion of indicators in the government focus set recovered by the system), an overlooked-indicator discovery rate (the proportion of relevant indicators outside the focus set that are identified), and a model--government focus ratio (comparing system and government coverage on a shared indicator subset). These quantities are then aggregated across episodes to produce corpus-level statistics\cite{8}.

\subsection{Results, case studies, and limitations}

Table~\ref{tab:system-comparison} summarizes aggregate results on the 1{,}027 policy episodes. Compared to two LLM-based baselines that predict impacted indicators directly from the policy text (GPT~5.1 and Doubao), PPCR-IM achieves the strongest performance on all three metrics. Its expected-indicator coverage score reaches a mean of 0.902 (std.\ 0.048), versus 0.851 and 0.803 for GPT~5.1 and Doubao, respectively. The overlooked-indicator discovery rate improves more markedly, with PPCR-IM reaching 0.603 (std.\ 0.092), compared to 0.352 and 0.291 for the baselines. The model--government focus ratio also increases from 1.098 and 1.023 to 1.356, indicating that PPCR-IM systematically broadens indicator coverage relative to what is emphasized in government focus sets while still recovering most expected indicators\cite{9}.

Qualitative inspection of individual episodes helps explain these gains. For fiscal transfer and tax policies, PPCR-IM not only recovers the intended macroeconomic indicators (such as growth and unemployment) but also highlights debt-related and distributional indicators that appear in the expert relevance sets but not in the government focus sets. The explicit DAG representation makes it possible to trace how such additional indicators arise through intermediate mechanisms, which the direct-prediction baselines cannot provide. At the same time, limitations remain: in some domains the generated consequence nodes are generic or weakly justified, and the indicator mapper sometimes assigns ambiguous directions when textual cues are scarce. These observations reinforce our positioning of PPCR-IM as a decision-support tool that produces auditable, structured hypotheses, rather than a system that can replace domain expertise in interpreting policy impacts\cite{10}.

\section{Demo Availability, Reproducibility}

\subsection{Online demo and interaction modes}

PPCR-IM is exposed through a web-based demo targeted at three main user groups: policy analysts and public administration researchers who want structured support for consequence reasoning; NLP and AI practitioners who want to study graph-based uses of LLMs; and students or educators who require concrete examples of automated policy-impact analysis. The demo front-end provides a simple form where users can paste a policy description, optionally specify basic context (jurisdiction, year, policy type), and select a configuration profile for the underlying model and graph-generation parameters\cite{13}.

After submission, the system runs the full pipeline and returns a multi-pane view. One pane displays the generated multi-layer DAG, with nodes grouped by depth and edges indicating hypothesized influence relations. A second pane lists the indicator-level assessments: for each of the nineteen indicators, the demo shows whether it is predicted to be affected, the qualitative direction (increase, decrease, ambiguous), and the supporting consequence nodes. A third pane summarizes the evaluation measures for that episode when reference annotations are available. Users can expand or collapse layers in the graph, inspect which nodes are linked to which indicators, and download the full JSON record for offline analysis.

\subsection{Code and reproduction}

We release the PPCR-IM implementation, together with configuration files and documentation, under an open-source Apache~2.0 license. The code, preprocessed dataset, and example configuration scripts are documented and linked via our project page at \url{https://n23pg3gbm5.coze.site}. The license allows non-commercial and commercial research use, adaptation, and redistribution, subject to attribution and preservation of the license terms. In practice, this means that research groups can integrate PPCR-IM into their own policy-analysis pipelines, extend the indicator vocabulary, or swap in different LLM backends, while public-sector teams can experiment with the system as part of internal analytic toolchains\cite{11}.

To reproduce the experiments in this paper, users can follow a three-step procedure: (i) install the Python environment and obtain access credentials for the chosen LLM backend; (ii) download the curated dataset of 1{,}027 policy episodes with annotations over nineteen World Bank–style indicators and place the XLSX file in the expected directory; and (iii) run the provided batch script with the same configuration used in our experiments, which generates JSON outputs and metric summaries. The repository also includes helper scripts to re-create Table~\ref{tab:system-comparison} and Figure~\ref{fig:region-pie}, and to run PPCR-IM side by side with the GPT~5.1 and Doubao baselines on new policy episodes\cite{12}.

\section{Conclusion}

In this paper, we introduced PPCR-IM, a system that turns qualitative public policy consequence analysis into multi-layer DAG construction with indicator alignment over a fixed vocabulary of nineteen World Bank–style social and macroeconomic indicators. On a curated dataset of 1{,}027 policy episodes with government focus sets and expert relevance annotations, PPCR-IM achieves a mean expected-indicator coverage score of 0.902 (std.\ 0.048), an overlooked-indicator discovery rate of 0.603 (std.\ 0.092), and a model--government focus ratio of 1.356, outperforming two strong LLM baselines (GPT~5.1 and Doubao) on all three metrics. These results indicate that making intermediate consequence structures explicit, rather than predicting indicators directly from text, helps recover most indicators emphasized in official documents while systematically surfacing additional relevant dimensions. Looking ahead, we plan to enrich the graph semantics with causal and uncertainty annotations, link DAGs to numerical simulation or historical outcome data, broaden the geographic and policy coverage of the corpus, and conduct user studies or expert workshops to better understand how PPCR-IM fits into real policy-analysis workflows.

\section{Ethics and Broder Impact Statement}
PPCR-IM is a decision-support system for structuring public-policy consequence reasoning; its outputs are hypothesis-level guidance, not causal predictions. It can improve transparency by externalizing multi-layer DAG pathways and mapping them to a shared indicator set, but LLM-generated nodes and qualitative directions may be biased, culturally contingent, or weakly supported and could mislead if treated as forecasts. We recommend human oversight, validation with evidence and stakeholders, and explicit uncertainty disclosure. The system requires no personal data; users must ensure inputs comply with privacy and governance rules.

\section{Limitations}
PPCR-IM relies on LLM-generated consequence DAGs, so nodes and links may be generic, incomplete, or biased. Indicator impacts are qualitative (increase/decrease/ambiguous) rather than quantitative, and directions can be uncertain when evidence is limited. The DAG represents plausible mechanisms but does not establish causal validity; outputs should be treated as structured hypotheses requiring expert verification. Finally, evaluation depends on government focus sets and expert annotations, which may reflect subjective or institutional priorities.

\bibliography{custom}

\end{document}